\documentclass[aps,prd,twocolumn,nofootinbib]{revtex4-1}
\usepackage{graphicx}
\usepackage{hyperref}
\usepackage{slashed}
\usepackage{color}

\begin{document}

\title {Excited doubly heavy baryon production via $W^+$ boson decays}

\author{Peng-Hui Zhang}
\email{zhangpenghui@cqu.edu.cn}
\author{Lei Guo}
\email{guoleicqu@cqu.edu.cn, correponding author}
\author{Xu-Chang Zheng}
\email{zhengxc@cqu.edu.cn}
\author{Qi-Wei Ke}
\email{keqw@cqu.edu.cn}

\address{College of Physics, Chongqing University, Chongqing 401331, P.R. China.}

\date{\today}

\begin{abstract}
In this paper, decay widths of the doubly heavy baryons ($\Xi_{cc} ~\text{and} ~\Xi_{bc}$) production are theoretically calculated in the whole phase space through $W^+ \to \Xi_{cc}+ \bar{c}+\bar{s}$ and $W^+ \to \Xi_{bc}+ \bar{b}+\bar{s}$, within the framework of nonrelativistic QCD (NRQCD). Differential widths $d\Gamma/ds_{12}$, $d\Gamma/ds_{23}$, $d\Gamma/dcos\theta_{12}$, and $d\Gamma/dcos\theta_{13}$ are also given. In addition to the ordinary S-wave contributions for the baryons, we specifically calculate P-wave contributions as a comparison, namely the high excited states of the intermediate diquark, including $[^1P_1]$ and $[^3P_J]$ (with $J=0,1,2$) in both color antitriplet state $\overline{\mathbf{3}}$ and color sextuplet state $\mathbf{6}$. It shows that the contribution from the P-wave is about one order lower than the S-wave. According to the results, we can expect plentiful events produced at the LHC, i.e., $3.69\times10^5$ $\Xi_{cc}$ events and $4.91\times10^4$ $\Xi_{bc}$ events per year.\\


\end{abstract}

\maketitle

\section{Introduction}

$W^+$ boson is one of the vector bosons that mediate the weak interaction, of which the decay properties are significant to the standard model (SM). Measurements about the width and the branching ratios of $W^+$ boson decay actually provide a way to determine the mixing of $c$ quark and $s$ quark ($|V_{cs}|$) \cite{DELPHI:1998hlc, PDG2020W}. Ascertainment for quark mixing matrix element $|V_{tb}|$ also involves the branching ratio $B(t \to W+b)$ \cite{D0:2010mwe, ParticleDataGroup:2018ovx}, and moreover, researches on $W$-physics can be a meaningful verification for the SM and a practicable access to new phenomena beyond the standard model \cite{CMS:2017zts, CMS:2017fgp, Cen:2018okf, CMS:2012bpt, Haisch:2018djm}.
 
Doubly heavy baryon is a notable prediction in the constituent quark model. It contains two heavy quarks, which can only be $cc$, $bc$, or $bb$, since top quark has already decayed before hadronization. Doubly heavy baryons with a strange quark include $\Omega^{+}_{cc}$, $\Omega^{0}_{bc}$, and $\Omega^{-}_{bb}$, while those with a light quark $u$ or $d$ include $\Xi^{++}_{cc}$, $\Xi^{+}_{cc}$, $\Xi^{+}_{bc}$, $\Xi^{0}_{bc}$, $\Xi^{0}_{bb}$, and $\Xi^{-}_{bb}$; we unify the related baryons as $\Xi_{cc}$ and $\Xi_{bc}$ for short in this paper. Searching for doubly heavy baryons can become a substantial test to the quark model, QCD theory, and gauge theory of strong interaction. Further studies also develop our comprehension of NRQCD theory, promote heavy flavor physics, and advance our understanding for the quark structure inside these heavy baryons \cite{Aliev:2020lly, LHCb:2021xba, LHCb:2020iko, Luchinsky:2020fdf}. 

Heavy quarkonia and $B_c$ meson have been discovered for a long time, and extensively investigated \cite{Brambilla:2010cs, Andronic:2015wma, Lansberg:2019adr, Chapon:2020heu, Bodwin:2013nua, CHCJXW08, Chang:2007si, CFQLPS11, Petrelli:1997ge, Zheng:2019egj, JJCFQ16, QLLXGW12}. Comparing to these, researches on doubly heavy baryons are not plentiful in both experiment and theory. Hitherto the unique observed doubly heavy baryon in experiment is $\Xi_{cc}^{++}$. It is first discovered in the channel $\Xi_{cc}^{++} \rightarrow \Lambda_{c}^{+} K^{-} \pi^{+} \pi^{+}$ and reported by the LHCb collaboration in 2017 \cite{LHCb17}, which is confirmed through its decay $\Xi_{cc}^{++} \to \Xi_c^+ \pi^+$ in following experiment \cite{LHCb:2018pcs}. Although there are no new doubly heavy baryons found yet, the discovery of $\Xi_{cc}^{++}$ certainly inspires people's great interest in further researches on them. As to the theoretical aspect, some papers concentrate on the direct production of the doubly heavy baryon at $e^+e^-$ colliders \cite{ZXCCHC16, JJXGW12}, the indirect production in the decay of Higgs or top quark \cite{Niu:2019xuq, JJN18}, and hardronic production \cite{Berezhnoy:1998aa, Zhang:2011hi, Chang:2006eu}. 

In the NRQCD framework \cite{Bodwin:1994jh}, the production of doubly heavy baryons can be divided into two procedures: the first one is that $W^+$ boson decays to produce four particles $c\bar{s}$ and $Q\bar{Q}$ ($Q$ for a heavy quark $c$ or $b$), and then $c$ and heavy quark $Q$ bind to a perturbatively heavy diquark $(Qc)[n]$ ($[n]$ is for color and spin state). The second procedure is that the hadronization of diquark $(Qc)[n]$ into the doubly heavy baryon $\Xi_{Qc}$, which is depicted by a nonperturbative factor. The nonperturbative factor can be related with the wave function at the origin $|\Psi(0)|$ for S-wave, or the derivative wave function at the origin $|\Psi^{\prime}(0)|$ for P-wave. $|\Psi(0)|$ and $|\Psi^{\prime}(0)|$ for heavy hadrons are derived from the experiment or some nonperturbative
methods, e.g., the potential model \cite{Kiselev:2000jc, Kiselev:2002iy}, lattice QCD (LQCD) \cite{Bodwin:1996tg}, or QCD sum rules \cite{Kiselev:1999sc}. It is necessary to mention here that the estimates for $|\Psi(0)|$ and $|\Psi^{\prime}(0)|$ have distinguishable disparities between different methods \cite{Buchmuller:1980su, Eich78, Quigg:1977dd, Martin:1980jx}. But fortunately, these nonperturbative elements are overall parameters, and cause changes of the final results only in the sense of constant times. To speak further, the essential part in the results, comparing to the overall factor $|\Psi(0)|$ and $|\Psi^{\prime}(0)|$, is the perturbative part from the first procedure $W^+ \to diquark$; dynamics of the perturbative diquark production is embodied in the shapes of the differential distributions.

In our paper, we shall analyse the indirect production of $\Xi_{cc}$ and $\Xi_{bc}$ via the main relevant channels of $W^+$-boson decay, i.e., $W^+ \to \Xi_{cc}+ \bar{c}+\bar{s}$ and $W^+ \to \Xi_{bc}+ \bar{b}+\bar{s}$. Due to the small value of $|V_{cb}|$ ($|\frac{V_{cb}}{V_{cs}}|^2 < 0.002$)
, the contribution from the channel $W^+ \to c{\bar{b}}$ is suppressed comparing with the channel $W^+ \to c{\bar{s}}$, so we do not consider $W^+ \to c{\bar{b}}$ in calculation. As the published date, the total width $\Gamma_W=2.085~\rm{GeV}$, and the branching ratio $B(W^+ \to cX)$ occupies $33.3\%$ in the $W^+$ decay mode \cite{PDG2020W}. With the collision energy $\sqrt{s}=14~\rm{TeV}$ and luminosity ${\cal L}=10^{34}~\rm{cm^{-2}\cdot s^{-1}}$, $W^+$ events are evaluated to be $3.07\times10^{10}$ per operation year at the LHC \cite{Gaunt:2010pi, Qiao:2011yk}. Here we can roughly estimate doubly heavy baryon events around $10^5$. Therefore, the LHC can generate sufficient doubly heavy baryons and offer a satisfactory platform to study $W$-physics. Apart from the hadronic collider, the proposed $e^+e^-$ colliders might provide less but specifical information as well, owing to the pure background, e.g., the Circular Electron Positron Collider (CEPC) and the International Linear Collider (ILC). The CEPC is designed to produce a total of $1\times10^{8}$ W bosons \cite{CEPCStudyGroup:2018ghi}; hence the doubly heavy baryons in the CEPC will be around order of $10^3$. Nevertheless, events from direct production at the CEPC can reach $10^5$ as well \cite{JJXGW12}.

The rest parts of this paper are organized as follows. In Sec. II, we make a review of the NRQCD formulation, and give the amplitudes for the process  $W^{+}\rightarrow \Xi_{Qc}+\bar{Q}+\bar{s}$. In Sec. III, the numeral results of the total widths and the derivative distributions are given in detail. Then we make a discussion on the uncertainties with different mass of quarks. The final section is reversed for a short summary.

\section{Calculation Techniques}

\begin{figure}[htb]
	\centering  
	\includegraphics[height=2.7cm,width=8cm]{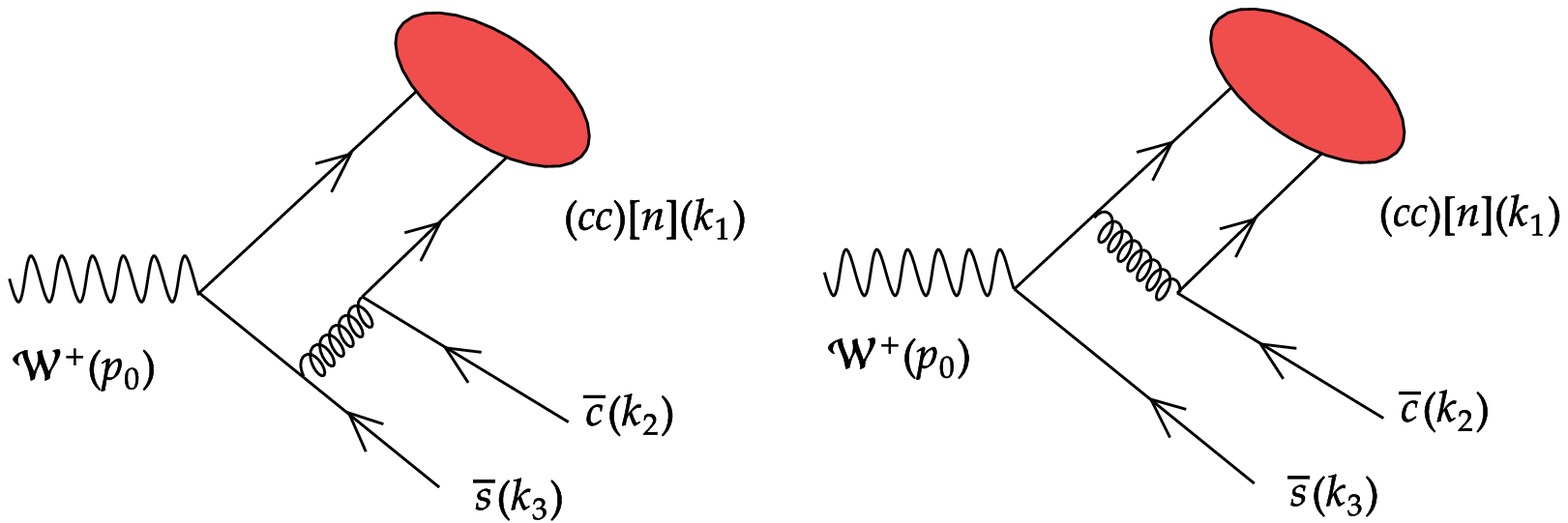}
	\includegraphics[height=2.7cm,width=8cm]{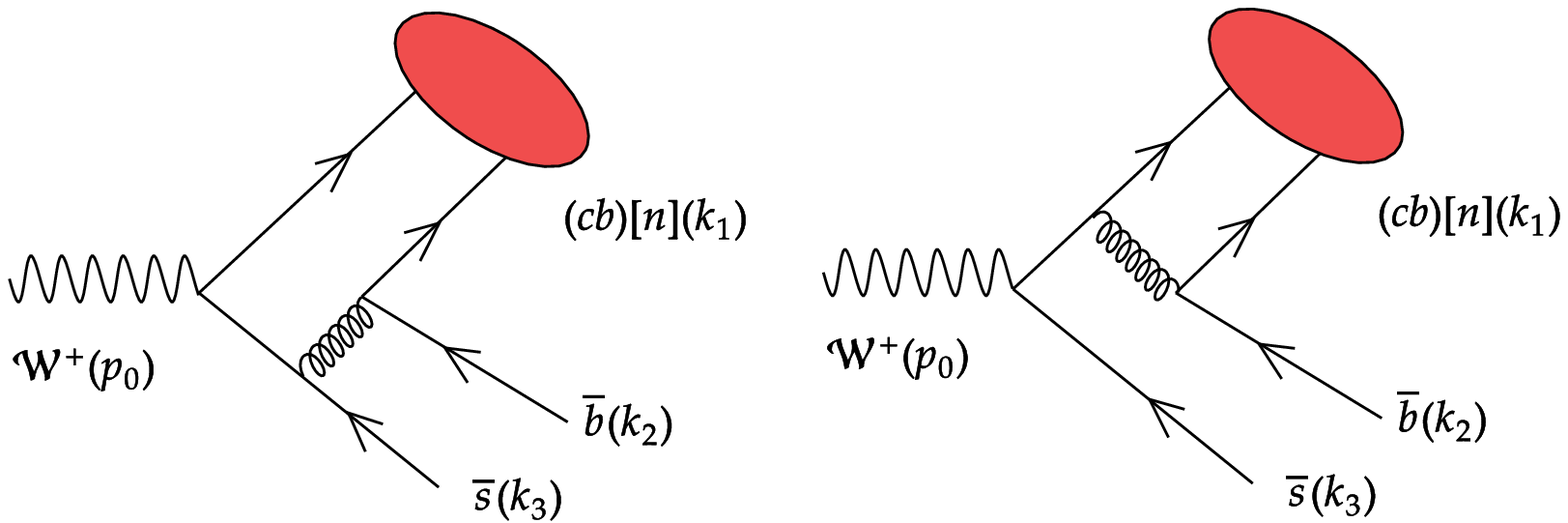} 
	\caption{Typical Feynman diagrams for the processes $W^+ (p_0)\rightarrow (cc)[n](k_1)+ \bar {c} (k_2) + \bar{s} (k_3)$ and  $W^+ (p_0)\rightarrow (bc)[n](k_1)+ \bar {b} (k_2) + \bar{s} (k_3)$, where $p_0$ and $k_i$ represent the four momenta of the associated particles.}  
	\label{feyn1}  
\end{figure}

As mentioned in section I, we shall deal with $W^+$ boson decays into the doubly heavy baryons $\Xi_{cc}$ and $\Xi_{bc}$, which can be written together as $\Xi_{Qc}$ for short. Fig.\ref{feyn1} shows the Feynman diagrams for the processes  $W^{+}\rightarrow (cc)[n]+\bar{c}+\bar{s}$ and  $W^{+}\rightarrow (bc)[n]+\bar{b}+\bar{s}$ respectively, where $[n]$ indicates the spin-color quantum number of the intermediate diquark state.

Under the framework of NRQCD effective theory \cite{Bodwin:1994jh}, the unpolarized differential decay width for $\Xi_{Qc}$ production through the channel $W^{+}(p_{0})\to \Xi_{Qc}(k_{1})+\bar{Q}(k_{2})+\bar{s}(k_{3})$ can be factorized as:
\begin{equation}\label{factorize}
	d\Gamma_{W^+ \to \Xi_{Qc}+ \bar{Q}+\bar{s}}=\sum_{n}{d\hat\Gamma_{W^+ \to (Qc)[n]+ \bar{Q}+\bar{s}} \langle{\cal O}^{\cal B}[n] \rangle},
\end{equation}
here $d\hat\Gamma_{W^+ \to (Qc)[n]+ \bar{Q}+\bar{s}}$ represents the differential decay width for $W^+ \to (Qc)[n]+ \bar{Q}+\bar{s}$, which is actually a short-distance coefficient of the hard process that $W^+$ decays into diquark state $(Qc)[n]$, and perturbatively calculable hence. The other factor $\langle{\cal O}^{\cal B}[n] \rangle$, namely the long-distance matrix element, characterizes the nonperturbative process from the diquark state $(Qc)[n]$ hadronizing to a doubly heavy baryon $\Xi_{Qc}$. $\cal B$ is short for any doubly heavy baryon $\Xi_{Qc}$.

The nonperturbative element $\langle{\cal O}^{\cal B}[n] \rangle$ is proportional to the transition probability of the diquark $(Qc)[n]$ hadronizing into $\Xi_{Qc}$, and can be obtained by calculating the origin value of wave function or its derivative, $\langle{\cal O}^{\cal B}[n] \rangle=|\Psi(0)|^2$ ($|\Psi^{\prime}(0)|^2$ for P-wave). Meanwhile, the perturbatively short-distance coefficient can be explicitly expressed as
\begin{equation}
	d\hat\Gamma_{W^+ \to (Qc)[n]+ \bar{Q}+\bar{s}}=\frac{1}{2 m_{W}} \overline{\sum}\left|M_{(Qc)[n]}\right|^{2} d \Phi_{3},
\end{equation} 
where $\overline{\sum}$ means to average over the spin states of the initial particles and to sum over the spin and color states of the final particles. $d\Phi_{3}$ is the three-body phase space, i.e.
\begin{equation}
	d{\Phi_3}=(2\pi)^4 \delta^{4}\left(p_0- \sum_i^3 k_{i}\right)\prod_{i=1}^3
	\frac{d^3 {k_i}}{(2\pi)^3 2 k_i^0}.
\end{equation}
By defining the invariant mass as $s_{ij}=(k_i+k_j)^2$, the differential decay width is equivalently written as
\begin{equation}
	d\hat\Gamma_{W^+ \to (Qc)[n]+ \bar{Q}+\bar{s}}=\frac{1}{256\pi^3 m_W^3}\overline{\sum}\left|M_{(Qc)[n]}\right|^{2}{ds_{12} ds_{23}}.
\end{equation}
The Feynman diagrams are generated by FeynArts 3.11 \cite{Hahn:2000kx}. With the help of FeynCalc 9.2 \cite{Shtabovenko:2020gxv} and three-body phase space formulas, we are able to calculate the total decay width, and moreover, the relevant differential distributions. More detailed formulas for the three-body phase space can be seen in Ref.\cite{Chang:2007si}.

\subsection{Amplitudes for the diquark production}

In the heavy diquark $(Qc)$ production, $W^+$ first produces two quarks $c\bar{s}$, and emits an intermediate gluon which is hard enough to generate a heavy quark pair $Q\bar{Q}$. Ergo, the amplitude for this process can be derived in the perturbative QCD (pQCD).

By the method of applying charge conjugation $C=$ $-i\gamma^{2} \gamma^{0}$, the hard amplitudes, corresponding to the perturbative parts in the baryons production, can be correlated with amplitudes of the quarkonium or meson production, which are more accustomed processes to us. It has been sufficiently demonstrated in Refs.\cite{ZXCCHC16, JJXGW12}, and here we are going to give a brief presentation. When dealing with the hard amplitudes for doubly heavy baryons, we should utilize charge conjugation to reverse one fermion line, generally writing as $L1=\bar{u}_{s_{1}}\left(k_{12}\right) \Gamma_{i+1} S_{F}\left(q_{i}, m_{i}\right) \cdots S_{F}\left(q_{1}, m_{1}\right) \Gamma_{1} v_{s_{2}}\left(k_{2}\right)$. Here $\Gamma_{i}$ is the interaction vertex, $S_{F}\left(q_{i}, m_{i}\right)$ is the fermion propagator, $s_{1}$ or $s_2$ is for spin index, and $i$ is the number of the fermion propagator ($i=0,1,...$) in this fermion line. This conversion obeys
\begin{eqnarray}
	v_{s_{2}}^{T}\left(k_{2}\right)C&=&-\bar{u}_{s_{2}}\left(k_{2}\right),\nonumber\\
	C^{-}\bar{u}_{s_{1}}\left(k_{12}\right)^{T}&=&v_{s_{1}}\left(k_{12}\right),\nonumber\\
	C^{-}S_{F}^{T}\left(-q_{i},m_{i}\right)C&=&S_{F}\left(q_{i}, m_{i}\right),\nonumber\\
	C^{-}\Gamma_{i}^{T}C&=&-\Gamma_{i}.
\end{eqnarray}
Inserting the identity $I=CC^{-}$, the fermion line $L1$ is reversed to
\begin{widetext}
	\begin{eqnarray}
		L1 = L1^T &=& v^T_{s_2}(k_2) \Gamma^T_1 F^T_F(q_1,m_1) \cdots S^T_F(q_{i},m_i) \Gamma^T_{i+1} \bar{u}^T_{s_{1}}\left(k_{12}\right) \nonumber\\
		&=& v^T_{s_2}(k_2) C C^- \Gamma^T_1 C C^- S^T_F(q_1,m_1) C C^- \cdots
		C C^- S^T_F(q_{i},m_i) C C^-\Gamma^T_{i+1} CC^- \bar{u}^T_{s_{1}} \left(k_{12}\right) \nonumber\\
		&=&(-1)^{(i+2)} \bar{u}_{s_2}(k_2)\Gamma_1 S_F(-q_1,m_1) \cdots S_F(-q_{i},m_i) \Gamma_{i+1} v_{s_{1}}\left(k_{12}\right).
	\end{eqnarray}

As an example, we first consider an amplitude of the process that $W^+$ boson decay to four free quarks, namely $W^+ \to c+b+\bar{b}+\bar{s}$ ($c$ and $b$ have not bound to diquark yet),
	\begin{eqnarray} 
		iM_{b1}&=&\frac{-i e V_{c s} {\cal C}}{sin{\theta}_W}\frac{-i}{\left(k_{12}+k_{2}\right)^{2}+i \epsilon} \bar{u}_{s1}\left(k_{12}\right)\left(i g_{s} \gamma^{\mu}\right) v_{s2}(k_2) \bar{u}_{s1'}(k_{11}) \slashed{\varepsilon}(p_{0}) P_L\nonumber \\ &&\frac{i}{-(\slashed{k}_{12}+\slashed{k}_{2}+\slashed{k}_{3})+i \epsilon} \left(i g_{s} \gamma_{\mu} \right) v_{s3}\left(k_{3}\right),
	\end{eqnarray}
where $k_{11}$ and $k_{12}$ denote the momenta of $c$ and $Q$ ($Q$ is $b$ quark for this diagram), ${\cal C}$ is the color factor, and $\theta_W$ is the Weinberg angle. After reversing the first fermion line, we obtain
	\begin{eqnarray} 
		iM_{b1}&=&(-1)^{i+2}\frac{-i e V_{c s} {\cal C}}{sin{\theta}_W}\frac{-i}{\left(k_{12}+k_{2}\right)^{2}+i \epsilon} \bar{u}_{s2}\left(k_{2}\right)\left(i g_{s} \gamma^{\mu}\right) v_{s1}(k_{12}) \bar{u}_{s1'}(k_{11}) \slashed{\varepsilon}(p_{0}) P_L\nonumber \\ &&\frac{i}{-(\slashed{k}_{12}+\slashed{k}_{2}+\slashed{k}_{3})+i \epsilon} \left(i g_{s} \gamma_{\mu} \right) v_{s3}\left(k_{3}\right).
	\end{eqnarray}
\end{widetext}
Here $\slashed{\varepsilon}(p_{0})$ is short for $\varepsilon_{\nu}(p_{0})\gamma^{\nu}$; $i=0$, since no fermion propagator exists in this fermion line. To form the diquark state, we suppose the relative velocity between two heavy quarks is small. More explicitly, we suppose that $k_{11}=\frac{m_c}{M_{Qc}}k_1+q$ and $k_{12}=\frac{m_Q}{M_{Qc}}k_1-q$, in which $k_1$ is the momentum of the diquark and $q$ is the small relative momentum between two components in the diquark. Meanwhile, $M_{Qc}\simeq m_Q+m_c$ is adopted in order to ensure the gauge invariance of the amplitude. Now we can insert the spin projector $\Pi$ and finally write the amplitude as
\begin{widetext}
	\begin{eqnarray} 
		iM_{b1}=\frac{-i e V_{c s} {\cal C}}{sin{\theta}_W}\frac{-i}{\left(k_{12}+k_{2}\right)^{2}+i \epsilon} \bar{u}_{s2}\left(k_{2}\right)\left(i g_{s} \gamma^{\mu}\right) \Pi \slashed{\varepsilon}(p_{0}) P_L\frac{i}{-(\slashed{k}_{12}+\slashed{k}_{2}+\slashed{k}_{3})+i \epsilon} \left(i g_{s} \gamma_{\mu} \right) v_{s3}\left(k_{3}\right).
	\end{eqnarray}
The projector $\Pi$ takes the form of
\begin{eqnarray}\label{spinprojector1}
	\Pi_{k_{1}}(q)=\frac{-\sqrt{M_{Qc}}}{4 m_{Q} m_{c}}\left(\slashed{k}_{12}-m_{Q}\right) \gamma^{5}\left(\slashed{k}_{11}+m_{c}\right), 
\end{eqnarray}
or
\begin{eqnarray}\label{spinprojector3}
	\Pi_{k_{1}}^{\beta}(q)=\frac{-\sqrt{M_{Qc}}}{4 m_{Q} m_{c}}\left(\slashed{k}_{12}-m_{Q}\right) \gamma^{\beta}\left(\slashed{k}_{11}+m_{c}\right),
\end{eqnarray}
\end{widetext}
for spin-singlet state or spin-triplet state respectively; we can equivalently adopt the simplified Eqs.(\ref{simplify1}, \ref{simplify2}). 
The S-wave amplitudes for the third picture in Fig.\ref{feyn1} can be written as
\begin{widetext}
	\begin{eqnarray}
		iM_{b1}[^1S_0]&=&\frac{-i e V_{c s} {\cal C}}{sin{\theta}_W}\frac{-i}{\left(k_{12}+k_{2}\right)^{2}+i \epsilon} \bar{u}_{s2}\left(k_{2}\right)\left(i g_{s} \gamma^{\mu}\right) \frac{-\sqrt{M_{bc}}}{4 m_{b} m_{c}}\left(\slashed{k}_{12}-m_{b}\right) \gamma^{5}\left(\slashed{k}_{11}+m_{c}\right) \slashed{\varepsilon}(p_{0}) P_L \nonumber \\&&\cdot \left.\frac{i}{-(\slashed{k}_{12}+\slashed{k}_{2}+\slashed{k}_{3})+i \epsilon} \left(i g_{s} \gamma_{\mu} \right) v_{s3}\left(k_{3}\right)\right|_{q=0}, \\
		iM_{b1}[^3S_1]&=&\varepsilon^s_{\beta}(k_1)\frac{-i e V_{c s} {\cal C}}{sin{\theta}_W} \frac{-i}{\left(k_{12}+k_{2}\right)^{2}+i \epsilon} \bar{u}_{s2}\left(k_{2}\right)\left(i g_{s} \gamma^{\mu}\right) \frac{-\sqrt{M_{bc}}}{4 m_{b} m_{c}}\left(\slashed{k}_{12}-m_{b}\right) \gamma^{\beta}\left(\slashed{k}_{11}+m_{c}\right) \slashed{\varepsilon}(p_{0}) P_L \nonumber \\&&\cdot \left.\frac{i}{-(\slashed{k}_{12}+\slashed{k}_{2}+\slashed{k}_{3})+i \epsilon} \left(i g_{s} \gamma_{\mu} \right) v_{s3}\left(k_{3}\right)\right|_{q=0}.
	\end{eqnarray}	

As to the P-wave amplitudes, the expression can be interconnected with the derivative of the S-wave expression in spin singlet or spin triplet respectively,
	\begin{eqnarray}
		iM_{b1}[^1P_1]&=&\varepsilon^l_{\alpha}(k_1)\frac{d}{dq_{\alpha}}\left[\frac{-i e V_{c s} {\cal C}}{sin{\theta}_W}\frac{-i}{\left(k_{12}+k_{2}\right)^{2}+i \epsilon}\right. \bar{u}_{s2}\left(k_{2}\right)\left(i g_{s} \gamma^{\mu}\right) \frac{-\sqrt{M_{bc}}}{4 m_{b} m_{c}}\left(\slashed{k}_{12}-m_{b}\right) \gamma^{5}\left(\slashed{k}_{11}+m_{c}\right) \slashed{\varepsilon}(p_{0})\nonumber \\&&\cdot \left.\left. P_L \frac{i}{-(\slashed{k}_{12}+\slashed{k}_{2}+\slashed{k}_{3})+i \epsilon} \left(i g_{s} \gamma_{\mu} \right) v_{s3}\left(k_{3}\right)\right]\right|_{q=0}, \\
		iM_{b1}[^3P_J]&=&\varepsilon^J_{\alpha\beta}(k_1)\frac{d}{dq_{\alpha}}\left[\frac{-i e V_{c s} {\cal C}}{sin{\theta}_W}\frac{-i}{\left(k_{12}+k_{2}\right)^{2}+i \epsilon}\right. \bar{u}_{s2}\left(k_{2}\right)\left(i g_{s} \gamma^{\mu}\right) \frac{-\sqrt{M_{bc}}}{4 m_{b} m_{c}}\left(\slashed{k}_{12}-m_{b}\right) \gamma^{\beta}\left(\slashed{k}_{11}+m_{c}\right) \slashed{\varepsilon}(p_{0}) \nonumber \\&&\cdot \left.\left. P_L\frac{i}{-(\slashed{k}_{12}+\slashed{k}_{2}+\slashed{k}_{3})+i \epsilon} \left(i g_{s} \gamma_{\mu} \right) v_{s3}\left(k_{3}\right)\right]\right|_{q=0},
	\end{eqnarray}	
\end{widetext}
where $\varepsilon^s_{\beta}(k_1)$ or $\varepsilon^l_{\alpha}(k_1)$ is the polarization vector that relates with the spin or orbit angular momentum of the diquark $(bc)$ in spin triplet S-state or spin singlet P-state; $\varepsilon^J_{\alpha\beta}(k_1)$ is the polarization tensor for the spin triplet P-wave states with $J$=0, 1 or 2. To select the appropriate total angular momentum, we perform polarization sum properly. The sum over polarization vector is
\begin{eqnarray}
	\sum_{r_{z}} \varepsilon^r_{\alpha} \varepsilon_{\alpha^{\prime}}^{r*}=\Pi_{\alpha \alpha^{\prime}},
\end{eqnarray}
$r$ stands for $s$ or $l$. The sum over polarization tensors are
\begin{eqnarray}
	\varepsilon_{\alpha \beta}^{0} \varepsilon_{\alpha^{\prime} \beta^{\prime}}^{0*} &=&\frac{1}{3} \Pi_{\alpha \beta} \Pi_{\alpha^{\prime} \beta^{\prime}}, \\
	\sum_{J_{z}} \varepsilon_{\alpha \beta}^{1} \varepsilon_{\alpha^{\prime} \beta^{\prime}}^{1 *} &=&\frac{1}{2}\left(\Pi_{\alpha \alpha^{\prime}} \Pi_{\beta \beta^{\prime}}-\Pi_{\alpha \beta^{\prime}} \Pi_{\alpha^{\prime} \beta}\right), \\
	\sum_{J_{z}} \varepsilon_{\alpha \beta}^{2} \varepsilon_{\alpha^{\prime} \beta^{\prime}}^{2 *} &=&\frac{1}{2}\left(\Pi_{\alpha \alpha^{\prime}} \Pi_{\beta \beta^{\prime}}+\Pi_{\alpha \beta^{\prime}} \Pi_{\alpha^{\prime} \beta}\right)-\frac{1}{3} \Pi_{\alpha \beta} \Pi_{\alpha^{\prime} \beta^{\prime}}.\nonumber
	\\
\end{eqnarray}
Here we define 
\begin{eqnarray}
	\Pi_{\alpha \beta}=-g_{\alpha \beta}+\frac{k_{1 \alpha} k_{1 \beta}}{M_{Qc}^{2}}.
\end{eqnarray}
Through the same approach, other amplitudes can be obtained. The derivatives of the spin projectors in the P-wave expressions can be simplified as Eqs.(\ref{simplify3}, \ref{simplify4}).

With respect to the color factors above, these have been illustrated in Refs.\cite{ZXCCHC16, JJXGW12, JJN18, Niu:2019xuq}. The color state for the diquark $(Qc)$ is either antitriplet $\overline{\mathbf{3}}$ or sextuplet $\mathbf{6}$ due to the decomposition of $SU_C(3)$ color group $\mathbf{3} \otimes \mathbf{3}=\overline{\mathbf{3}} \oplus \mathbf{6}$. The color factor is $\mathcal{C}_{i j, k}=\mathcal{N} \times \sum_{a, m, n}\left(T^{a}\right)_{i m}\left(T^{a}\right)_{j n} \times G_{m n k}$; $G_{m n k}$ is the antisymmetric function $\varepsilon_{mnk}$ for color antitriplet or the symmetric function $f_{mnk}$ for color sextuplet, and $\mathcal{N}=1 /\sqrt{2}$ is the normalization constant. Finally, in the squared amplitudes, we acquire that $\mathcal{C}^2$ is $\frac{4}{3}$ for the color antitriplet state, and $\frac{2}{3}$ for the color sextuplet state. In consideration of exchange antisymmetry from the identical quarks, the quantum number of the diquark $(cc)$ is $[^1S_0]_{{\mathbf{6}}}$, $[^3S_1]_{\overline{\mathbf{3}}}$, $[^1P_1]_{\overline{\mathbf{3}}}$, or $[^3P_J]_{\mathbf{6}}$, while the diquark $(bc)$ allows all color and spin states, i.e., $[^1S_0]_{\overline{\mathbf{3}}}$, $[^1S_0]_{{\mathbf{6}}}$, $[^3S_1]_{\overline{\mathbf{3}}}$, $[^3S_1]_{{\mathbf{6}}}$, $[^1P_1]_{\overline{\mathbf{3}}}$, $[^1P_1]_{{\mathbf{6}}}$, $[^3P_J]_{\overline{\mathbf{3}}}$, and $[^3P_J]_{{\mathbf{6}}}$.

\subsection{Hadronization}

The hadronization of the diquark into doubly heavy baryon is nonperturbative; the effect of this procedure is extracted into an overall coefficient $\langle{\cal O}^{\cal B}\rangle$ Eq.(\ref{factorize}), which has been associated with the wave function at the origin. We shall not distinguish the wave function for different color states $\overline{\mathbf{3}}$ and ${\mathbf{6}}$ as Refs.\cite{Petrelli:1997ge, Ma:2003zk, JJXGW12, Niu:2019xuq, JJN18}. Some people argue that the interaction inside the diquark with $\overline{\mathbf{3}}$ state is attractive while repulsive for the diquark with $\mathbf{6}$ state owing to the one-gluon exchange interaction, so $\mathbf{6}$ is suppressed to $\overline{\mathbf{3}}$ by order $v^2$ and its contribution can be ignored \cite{ZXCCHC16, Ma:2003zk}. But another view is that $\mathbf{6}$ and $\overline{\mathbf{3}}$ are of the same importance \cite{JJXGW12, JJN18, Niu:2019xuq, Ma:2003zk}. 

We use $h_{\overline{\mathbf{3}}}$ and $h_{\mathbf{6}}$ to represent the transition probabilities of the color antitriplet state and the color sextuplet state. According to NRQCD, $\Xi_{Qc}$, which is a bound state of two heavy
quarks with other light dynamical freedoms of QCD, can be expanded to a series of Fock states,
\begin{eqnarray}
	\left|\Xi_{Qc}\right\rangle=c_{1}(v)\left|\left(Qc\right) q\right\rangle+c_{2}(v)\left|\left(Qc\right) q g\right\rangle+\nonumber\\
	c_{3}(v)\left|\left(Qc\right) q g g\right\rangle+\cdots,
\end{eqnarray}
where $v$ is a small relative velocity between heavy quarks in the rest frame of the diquark. For a diquark in $\overline{\mathbf{3}}$ state, one of the heavy quarks can emit a gluon without changing the spin of the heavy quark, and this gluon then splits to a quark pair $q\bar{q}$. The heavy diquark can catch the light quark $q$ to form the baryon. As for $\mathbf{6}$ state, if the baryon is formed by $\left|\left(Qc\right) q \right\rangle$, the emitted gluon must change the spin of the heavy quark, leading to a suppression to
 $h_{\mathbf{6}}$. But it can formed from the component $\left|\left(Qc\right) q g\right\rangle$ as well. One of the heavy quarks emits a gluon without changing the spin of the heavy quark, and this gluon splits into $q\bar{q}$. The light quarks can also emit gluons, then the component can be formed with $qg$. Since a light quark can emit gluons easily, these contributions are at the same level, i.e., $c_{1}(v) \sim c_{2}(v) \sim c_{3}(v)$ \cite{Ma:2003zk}. Then we can take the assumption that
\begin{eqnarray}
	h_{6} \simeq h_{\overline{3}}&=&\left\langle\mathcal{O}^{\cal B}\right\rangle=\left|\Psi(0)\right|^{2}~( \text{or}~\left|\Psi^{\prime}(0)\right|^{2}).
\end{eqnarray}
The wave function at the origin can naturally connect with the radial wave function at the origin,
\begin{eqnarray}
|\Psi(0)|^2&=&\frac{1}{4\pi}|R(0)|^2,\nonumber\\
|\Psi^{\prime}(0)|^2&=&\frac{3}{4\pi}|R^{\prime}(0)|^2.
\end{eqnarray}

\section{Numerical Results}
The input parameters are adopted as:
\begin{eqnarray}
		&m_{c}=1.8~\mathrm{GeV}, ~m_{b}=5.1~\mathrm{GeV},\nonumber\\
		&m_{W}=80.4~\mathrm{GeV}, ~m_{Z}=91.2~\mathrm{GeV},\nonumber \\
		&e=\sqrt{\displaystyle{\frac{4\pi}{128}}}, ~cos(\theta_W)=\displaystyle{\frac{m_W}{m_Z}},~\left|V_{c s}\right|=1.
\end{eqnarray}
As a common choice, here masses are kept the same with Refs.\cite{Baranov:1995rc, Chang:2006eu}. We will use the same values of $|R(0)|$ and $|R^{\prime}(0)|$ as Ref.\cite{Kiselev:2002iy} in our paper, which are calculated in the $K^2O$ potential motivated by QCD with a three-loop function,
\begin{eqnarray}
	&|R_{cc}(0)|=0.523~\mathrm{GeV^{\frac{3}{2}}}, |R_{cc}^{\prime}(0)|=0.102~\mathrm{GeV^{\frac{5}{2}}},\nonumber\\
	&|R_{bc}(0)|=0.722~\mathrm{GeV^{\frac{3}{2}}}, |R_{bc}^{\prime}(0)|=0.200~\mathrm{GeV^{\frac{5}{2}}}.
\end{eqnarray}
The renormalization scale is set to be $2m_c$. From the solution of the five-loop renormalization group equation, we obtain $\alpha_s(2m_c)=0.239$ \cite{JJN18, Baikov:2016tgj, Herzog:2017ohr}. Since the mass of $s$ quark is so small, we will take $m_s=0$ in our calculations and it only causes a difference to the total value less than $10^{-5}$ order of magnitude.

\subsection{$\Xi_{cc}$ baryons}

We have listed the decay widths of $W^+ \to cc[n]+{\bar{c}}+\bar{s}$ in Table \ref{ccwidth}. Here the branching ratio is defined as
\begin{eqnarray}
	Br[n]=\frac{\Gamma_{W^+ \to \Xi_{cc}[n]+ \bar{c}+\bar{s}}}{\Gamma_W},
\end{eqnarray}
$[n]$ is the intermediate spin-color state. S-wave stands for the sum of states $[^1S_0]_{\mathbf{6}}$ and $[^3S_1]_{\overline{\mathbf{3}}}$. P-wave stands for the sum of states $[^1P_1]_{\overline{\mathbf{3}}}$ and $[^3P_J]_{\mathbf{6}}$ ($J=0,1,2$). We find that
\begin{itemize}
	\item The contribution from $[^3S_1]_{\overline{\mathbf{3}}}$ is the largest one among these states. The decay width of $[^1S_0]_{\mathbf{6}}$ is about $48\%$ of that of $[^3S_1]_{\overline{\mathbf{3}}}$. For S-wave, the color state is antitriplet for spin triplet and sextuplet for spin singlet; the former color factor is twice of the latter one.
	\item  The total decay width of P-wave is about one order lower than that of S-wave. The decay widths from $[^1P_1]_{\overline{\mathbf{3}}}$, $[^3P_0]_{\mathbf{6}}$, $[^3P_1]_{\mathbf{6}}$, and $[^3P_2]_{\mathbf{6}}$ are about $3.5\%$, $2.6\%$, $2.9\%$, and $1.1\%$ of that from $[^3S_1]_{\overline{\mathbf{3}}}$, respectively. Comparing with S-wave states, contributions from P-wave states are considerable when detailed calculations are needed.
	\item At the LHC, there are totally $3.69\times10^5$ $\Xi_{cc}$ events from $W^+$ decays per year, which indicates that the LHC can be a fruitful platform for doubly charmed baryon researches. Events from P-wave
	 states are of considerable quantity as well, reaching $10^4$-order. 
\end{itemize}

\begin{table}[htb]
\begin{tabular}{|c||c||c||c|}
\hline
~~State~~&~Decay width~&~Branching ratio~&~Events~ \\
\hline\hline
$[^1S_0]_{\mathbf{6}}$     & 7.64 & $3.67\times10^{-6}$ & ~$1.13\times10^5$~ \\
\hline
$[^3S_1]_{\overline{\mathbf{3}}}$     & 15.8 & $7.59\times10^{-6}$ & $2.33\times10^5$ \\
\hline
$[^1P_1]_{\overline{\mathbf{3}}}$     & 0.561 & $2.69\times10^{-7}$ & $8.26\times10^3$ \\
\hline
$[^3P_0]_{\mathbf{6}}$     & 0.404 & $1.94\times10^{-7}$ & $5.96\times10^3$ \\
\hline
$[^3P_1]_{\mathbf{6}}$     & 0.455 & $2.18\times10^{-7}$ & $6.70\times10^3$ \\
\hline
$[^3P_2]_{\mathbf{6}}$     & 0.169 & $8.09\times10^{-8}$ & $2.48\times10^3$ \\
\hline
S-wave    & 23.5 & $1.13\times10^{-5}$ & $3.46\times10^5$ \\
\hline
P-wave     & 1.59 & $7.62\times10^{-7}$ & $2.34\times10^4$ \\
\hline
Total    & 25.1 & $1.20\times10^{-5}$ & $3.69\times10^5$ \\
\hline
\end{tabular}
\caption{Decay widths (in unit: keV), branching ratios, and events at the LHC for the production of $\Xi_{cc}$ via $W^+$ decays. States represent the spin and color states for the intermediate diquark.}
\label{ccwidth}
\end{table}

\begin{figure}[h]
	\includegraphics[width=0.39\textwidth]{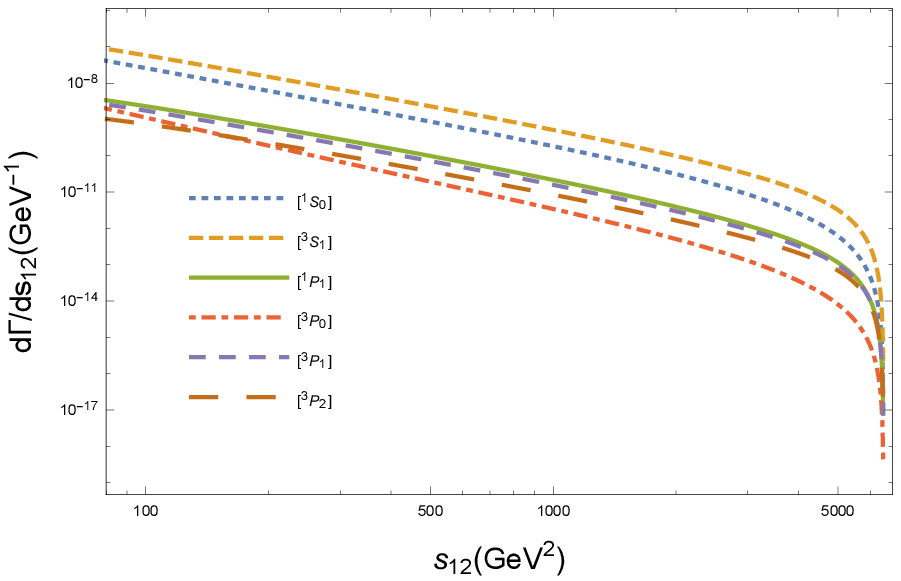}
	\includegraphics[width=0.39\textwidth]{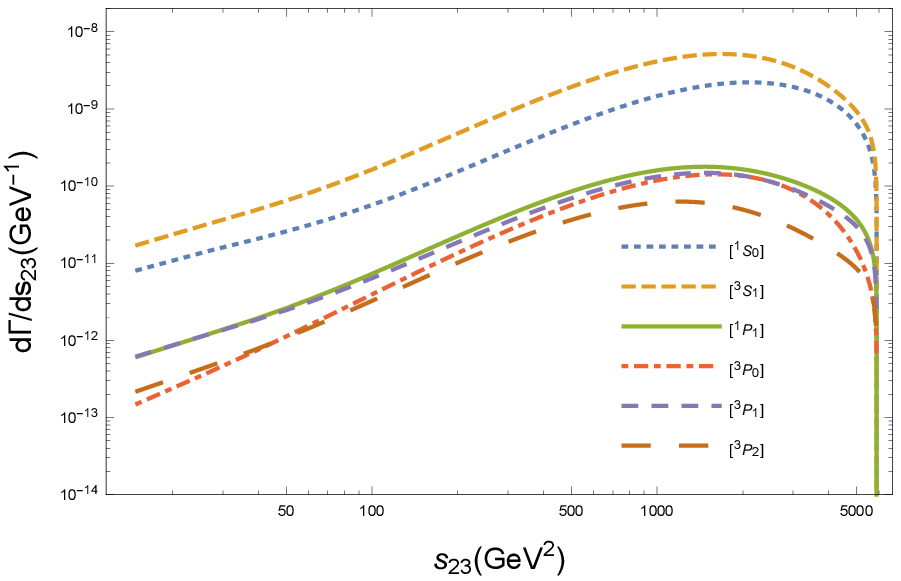}
	\caption{The invariant mass differential decay widths $d\Gamma/ds_{12}$ (top) and $d\Gamma/ds_{23}$ (bottom) for $W^+ \to \Xi_{cc}[n]+ \bar{c}+\bar{s}$, where $[n]$ stands for different state of the intermediate diquark. The subscripts ``1, 2, 3" denote $\Xi_{cc}$, $\bar{c}$, and $\bar{s}$ in sequence.} \label{ccs}
\end{figure}
\begin{figure}[h]
	\includegraphics[width=0.39\textwidth]{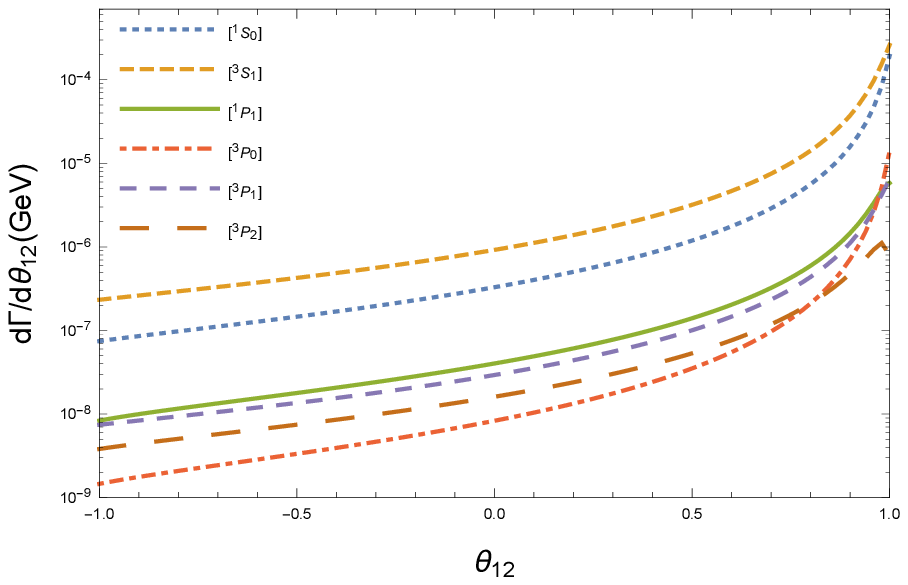}
	\includegraphics[width=0.39\textwidth]{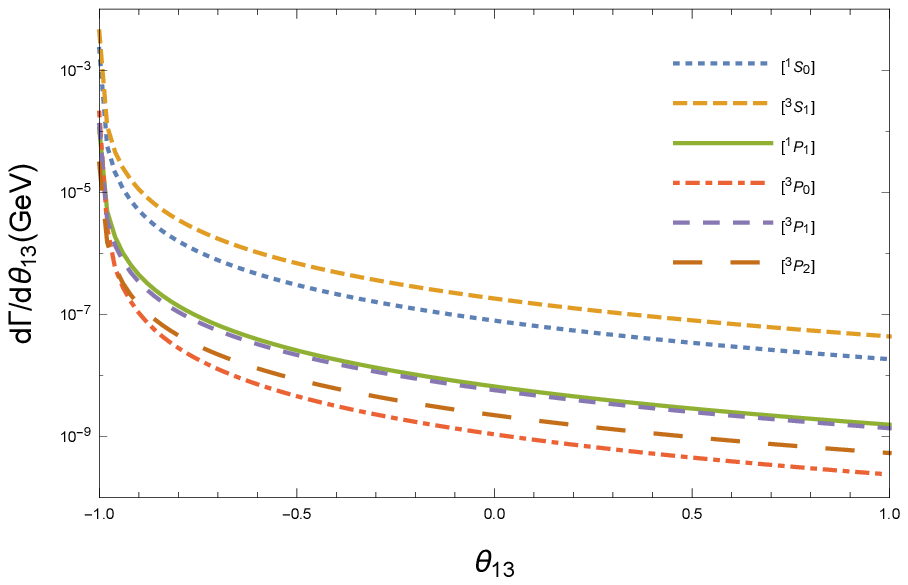}
	\caption{The angular differential decay widths $d\Gamma/dcos\theta_{12}$ (top) and $d\Gamma/dcos\theta_{23}$ (bottom) for $W^+ \to \Xi_{cc}[n]+ \bar{c}+\bar{s}$, where $[n]$ stands for different state of the intermediate diquark. The subscripts ``1, 2, 3" denote $\Xi_{cc}$, $\bar{c}$, and $\bar{s}$ in sequence.} \label{ccth}
\end{figure}

In order to show the characteristics of the decay $W^+ \to \Xi_{cc}[n]+{\bar{c}}+\bar{s}$, we derive the differential distributions of the invariant masses $s_{12}$ and $s_{23}$ in Fig.\ref{ccs}, as well as the differential distributions of the angles $\theta_{12}$ and $\theta_{13}$ in Fig.\ref{ccth}. Here $\theta_{ij}$ means the angle between outgoing three momenta $\overrightarrow{k_i}$ and $\overrightarrow{k_j}$ in the rest frame of $W^+$ boson. We use different size of dotted lines to represent the differential distributions from six intermediate states. Since one spin state has one corresponding color state for $\Xi_{cc}$ baryon, we simply tag lines with spin states in these figures, e.g., $[^1S_0]$ and $[^1P_1]$ mean $[^1S_0]_{\mathbf{6}}$ and $[^1P_1]_{\overline{\mathbf{3}}}$ respectively.

Figure.\ref{ccs} shows that the behaviors for the productions from the different states are similar to each other. In Fig.\ref{ccth}, we can find that the angular differential width $d\Gamma/dcos\theta_{12}$ takes its maximum value when $\theta_{12}=0$, i.e., the $\Xi_{cc}$ baryon and $\bar{c}$ quark move side by side in the $W^+$ rest frame; however, $d\Gamma/dcos\theta_{13}$ takes its maximum value when $\theta_{13}=\pi$, i.e., the $\Xi_{cc}$ baryon and $\bar{s}$ quark move back to back.

\subsection{$\Xi_{bc}$ baryons}

The production of $\Xi_{bc}$ baryon through $W^+ \to cb[n]+{\bar{b}}+\bar{s}$ is similar to the $\Xi_{cc}$ baryon production. The widths, branching ratios, and events at the LHC, are listed in Table \ref{cbwidth}, and we find that
\begin{table}[h]
	\begin{tabular}{|c||c||c||c|}
		\hline
		~~State~~&~Decay width~&~Branching ratio~&~~~~Events~~~~ \\
		\hline\hline
		$[^1S_0]_{\overline{\mathbf{3}}}$    & 1.095 & $5.25\times10^{-7}$ & $1.61\times10^{4}$ \\
		\hline
		$[^3S_1]_{\overline{\mathbf{3}}}$    & 0.958 & $4.59\times10^{-7}$ & $1.41\times10^{4}$ \\
		\hline
		$[^1S_0]_{\mathbf{6}}$    & 0.548 & $2.63\times10^{-7}$ & $8.06\times10^{3}$ \\
		\hline
		$[^3S_1]_{\mathbf{6}}$    & 0.479 & $2.30\times10^{-7}$ & $7.05\times10^{3}$ \\
		\hline
		$[^1P_1]_{\overline{\mathbf{3}}}$     & 0.030 & $1.43\times10^{-8}$ & $4.37\times10^{2}$ \\
		\hline
		$[^3P_0]_{\overline{\mathbf{3}}}$     & 0.078 & $3.76\times10^{-8}$ & $1.15\times10^{3}$ \\
		\hline
		$[^3P_1]_{\overline{\mathbf{3}}}$     & 0.057 & $2.74\times10^{-8}$ & $8.43\times10^{2}$ \\
		\hline
		$[^3P_2]_{\overline{\mathbf{3}}}$     & 0.0047 & $2.27\times10^{-9}$ &$6.97\times10$ \\
		\hline
		$[^1P_1]_{\mathbf{6}}$     & 0.015 & $7.13\times10^{-9}$ & $2.19\times10^{2}$ \\
		\hline
		$[^3P_0]_{\mathbf{6}}$     & 0.039 & $1.88\times10^{-8}$ & $5.77\times10^{2}$ \\
		\hline
		$[^3P_1]_{\mathbf{6}}$     & 0.029 & $1.37\times10^{-8}$ & $4.21\times10^{2}$ \\
		\hline
		$[^3P_2]_{\mathbf{6}}$     & 0.0024 & $1.14\times10^{-9}$ & $3.49\times10$ \\
		\hline
		S-wave    & 3.08 & $1.48\times10^{-6}$ & $4.53\times10^4$ \\
		\hline
		P-wave     & 0.255 & $1.22\times10^{-7}$ & $3.76\times10^3$ \\
		\hline
		Total    & 3.33 & $1.60\times10^{-6}$ & $4.91\times10^4$ \\
		\hline
	\end{tabular}
	\caption{Decay widths (in unit: keV), branching ratios, and events at the LHC for $\Xi_{bc}$ production via $W^+$ decays. States represent the spin-color states of the intermediate diquark.}
	\label{cbwidth}
\end{table}
\begin{figure}[h]
	\includegraphics[width=0.39\textwidth]{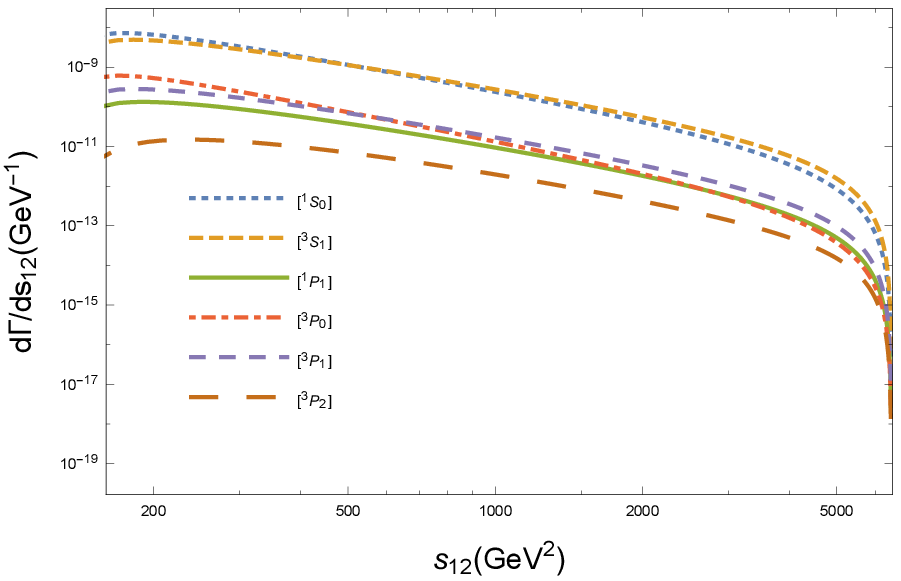}
	\includegraphics[width=0.39\textwidth]{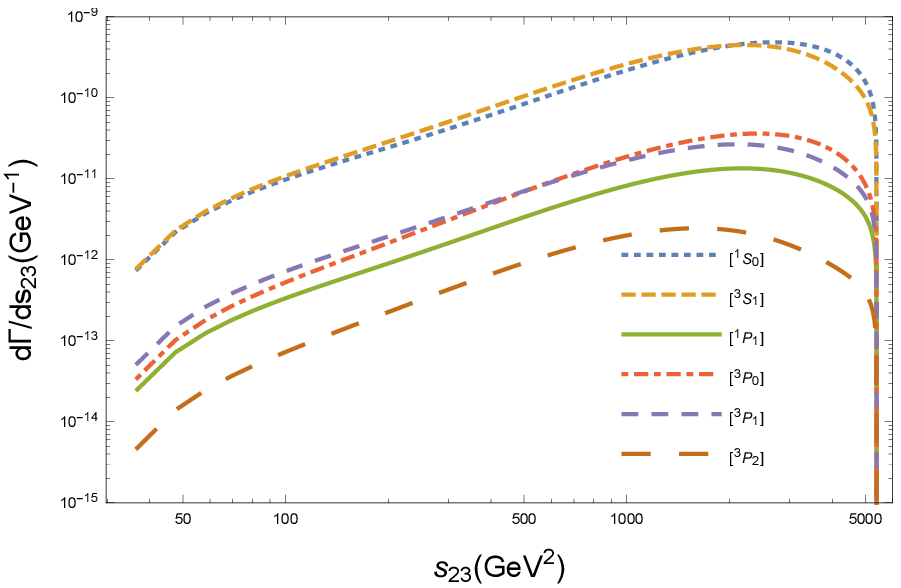}
	\caption{The invariant mass differential decay widths $d\Gamma/ds_{12}$ (top) and $d\Gamma/ds_{23}$ (bottom) for $W^+ \to \Xi_{bc}[n]+ \bar{b}+\bar{s}$, where $[n]$ stands for different state of the intermediate diquark. The subscripts ``1, 2, 3" denote $\Xi_{bc}$, $\bar{b}$, and $\bar{s}$ in sequence. } \label{cbs}
\end{figure}
\begin{figure}[h]
	\includegraphics[width=0.39\textwidth]{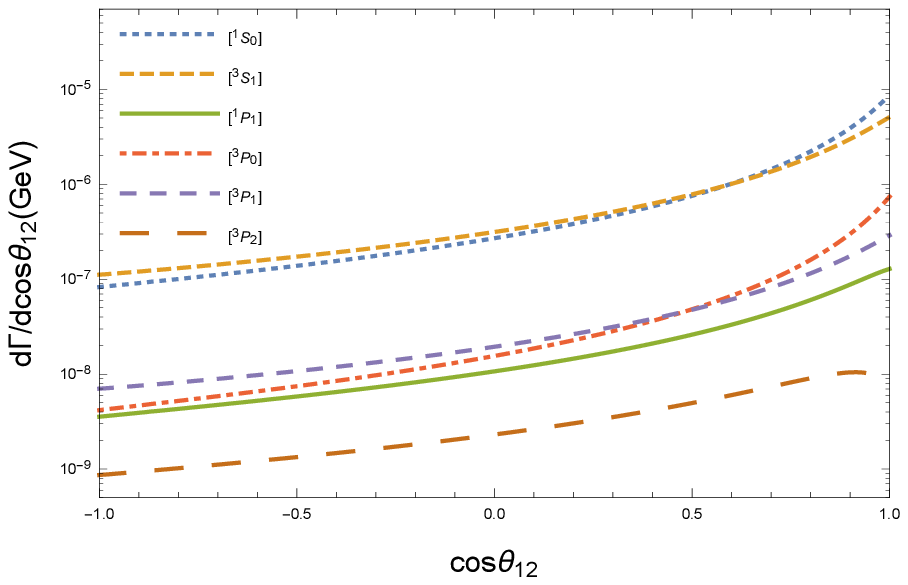}
	\includegraphics[width=0.39\textwidth]{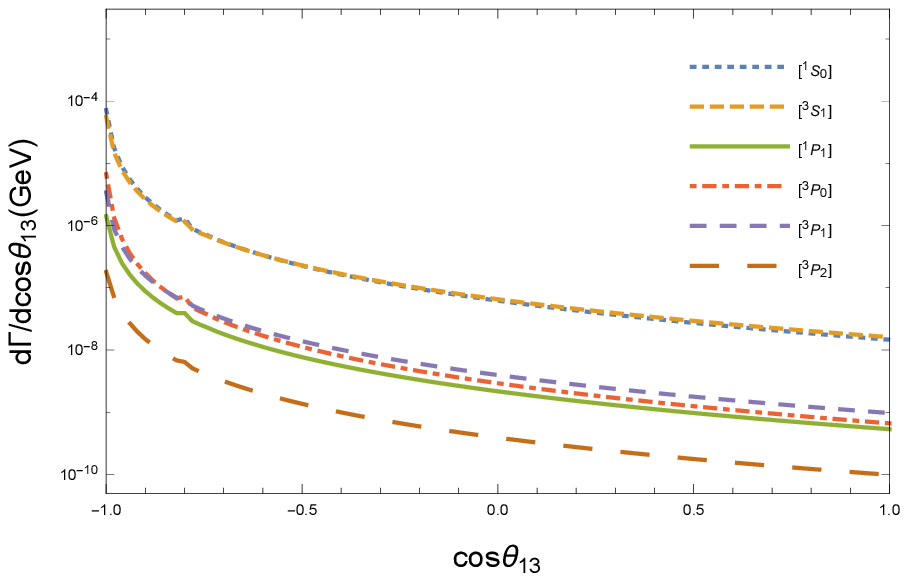}
	\caption{The angular differential decay widths $d\Gamma/dcos\theta_{12}$ (top) and $d\Gamma/dcos\theta_{23}$ (bottom) for $W^+ \to \Xi_{bc}[n]+ \bar{b}+\bar{s}$, where $[n]$ stands for different state of the intermediate diquark. The subscripts ``1, 2, 3" denote $\Xi_{bc}$, $\bar{b}$, and $\bar{s}$ in sequence. } \label{cbth}
\end{figure}

\begin{itemize}
	\item Comparing with $\Xi_{cc}$ production, the decay widths for $\Xi_{bc}$ are lower by about one order of magnitude. This can be understood from the production process. As shown in Fig.\ref{feyn1}, the $W^+$ boson decays into $c$ quark and $\bar{s}$ quark, then emits a hard gluon that generates a heavy quark pair $c\bar{c}$ or $b\bar{b}$. It is more difficult for the hard gluon to generate $b$-quark-pair than $c$-quark-pair, and the production of $\Xi_{bc}$ is consequently suppressed.
	\item The biggest contribution is from $[^1S_0]_{\overline{\mathbf{3}}}$. By adding up the same color states, the decay widths for spin states $[^3S_1]$, $[^1P_1]$, $[^3P_0]$, $[^3P_1]$, and $[^3P_2]$ are about $87.5\%$, $2.7\%$, $7.1\%$, $5.2\%$, $0.43\%$ of that for $[^1S_0]$.
	\item At the LHC, there are totally $4.91\times10^4$ $\Xi_{bc}$ events per year, which include $4.53\times10^4$ events coming from S-wave states and $3.76\times10^3$ events coming from P-wave states. If considering the LHC possible update with the higher luminosity ${\cal L}=10^{36}~\rm{cm^{-2}\cdot s^{-1}}$, these events can increase again by two order of magnitudes.
\end{itemize}

We present the differential widths of the invariant masses and the angles in Figs.\ref{cbs},\ref{cbth} to show the behaviors of the decay process. For $\Xi_{bc}$ baryon, both color antitriplet state and color sextuplet state are allowed for any spin states, so there are twelve states in total. To make these figures clear to see, we add the same color states up for different spin states, e.g., the line labeled with $[^1S_0]$ means the sum of contributions from $[^1S_0]_{\mathbf{6}}$ and $[^1S_0]_{\overline{\mathbf{3}}}$. These figures seem alike to Figs.\ref{ccs},\ref{ccth}. Those similarities of the angular and invariant mass differential widths also indicate the similar kinematic behaviors between the $\Xi_{cc}$ and $\Xi_{bc}$ productions in $W^+$ decays.
	
\subsection{Uncertainty analysis}
We have mentioned that only the channel $W^+ \to {c}+\bar{s}$ is accounted in our calculation. In fact, production of doubly heavy hadron in $W^+$ decays might have two channels, i.e., $W^+ \to {c}+\bar{s}$ and $W^+ \to {c}+\bar{b}$. For charmonium, the contribution of $W^+ \to c{\bar{c}}+c+\bar{b}$ is suppressed by three orders comparing with $W^+ \to c{\bar{c}}+c+\bar{s}$, due to the small value of $|V_{cb}|$. Therefore, theoretical estimates can ignore it. In the case of $B_c$ meson production, the contribution of $W^+ \to c{\bar{b}}+b+\bar{b}$ is also suppressed comparing with the contribution of $W^+ \to c{\bar{b}}+b+\bar{s}$. However, except $W^+ \to c{\bar{b}}+b+\bar{b}$, there is a second way to form $B_c$ meson in $W^+ \to {c}+\bar{b}$ channel, i.e., through $W^+ \to c{\bar{b}}+c+\bar{c}$ the antibottom quark directly from $W^+$ combines with the charmed quark from the intermediate gluon. Since the intermediate gluon is much easier to generate $c\bar{c}$ quark pair than $b\bar{b}$ quark pair, the width of $W^+ \to c{\bar{b}}+c+\bar{c}$ is only one order less than the width of $W^+ \to c{\bar{b}}+b+\bar{s}$ \cite{QLLXGW12}. This contribution is comparable with the P-wave contribution from the channel $W^+ \to c{\bar{s}}$; in that case the channel $W^+ \to c{\bar{b}}$ is non-negligible for $B_c$ meson. But fortunately this discussion does not exist in the doubly heavy baryon production. There is no such way to form $\Xi_{bc}$ baryon from $W^+ \to c\bar{b}$. The heavy quark $b$ in $\Xi_{bc}$ can only be provided by the intermediate gluon, and the antibottom quark directly from $W^+$ boson is a free quark. As for the process $W^+ \to cb+{\bar{b}}+\bar{b}$, it is three orders of magnitude smaller than $W^+ \to cb+{\bar{b}}+\bar{s}$ as above. Finally, we can directly ignore the channel $W^+ \to {c}+\bar{b}$.

The transition probability and the strong coupling constant have apparent theoretical uncertainties, but they influence the results merely as an overall factor, so here we does not discuss them. Apart from them, the decay width is also sensitive to quark masses $m_c$ and $m_b$. We vary $m_c=1.8\pm0.3~\mathrm{GeV}$ for $\Xi_{cc}$ production and $m_b=5.1\pm0.3~\mathrm{GeV}$ for $\Xi_{bc}$ production to obtain the uncertainties, which are presented in Tables (\ref{ccunc}, \ref{cbunb}) respectively. In Table \ref{cbunb}, state $[n]$ means the sum of the results from $[n]_{\overline{\mathbf{3}}}$ and $[n]_{\mathbf{6}}$.
\begin{table}[htb]
	\begin{tabular}{|c||c||c||c|}
		\hline
		~~State~~&~$m_c=1.5~$~&~$m_c=1.8$~&~$m_c=2.1$~ \\
		\hline\hline
		$[^1S_0]_{\mathbf{6}}$   & 13.38 & 7.64 & 4.75 \\
		\hline
		$[^3S_1]_{\overline{\mathbf{3}}}$  & 27.66  & 15.8 & 9.83 \\
		\hline
		$[^1P_1]_{\overline{\mathbf{3}}}$   & 1.42 & 0.561 & 0.254 \\
		\hline
		$[^3P_0]_{\mathbf{6}}$  & 1.01  & 0.404 & 0.185 \\
		\hline
		$[^3P_1]_{\mathbf{6}}$   & 1.15 & 0.455 & 0.208 \\
		\hline
		$[^3P_2]_{\mathbf{6}}$   & 0.427 & 0.169 & 0.077 \\
		\hline
		Total  & 45.06 & 25.1 & 15.3 \\
		\hline
	\end{tabular}
	\caption{Decay widths (in unit: keV) for the production of $\Xi_{cc}$ via $W^+$ decays by varying $m_c$ (in unit: GeV).}
	\label{ccunc}
\end{table}

\begin{table}[htb]
	\begin{tabular}{|c||c||c||c|}
		\hline
		~~State~~&~$m_b=4.8~$~&~$m_b=5.1$~&~$m_c=5.4$~ \\
		\hline\hline
		$[^1S_0]$   & 2.00 & 1.64 & 1.36 \\
		\hline
		$[^3S_1]$  & 1.76  & 1.44 & 1.18 \\
		\hline
		$[^1P_1]$   & 0.0566 & 0.0446 & 0.0356 \\
		\hline
		$[^3P_0]$  & 0.145 & 0.118 & 0.0962 \\
		\hline
		$[^3P_1]$   & 0.108 & 0.0858 & 0.0690 \\
		\hline
		$[^3P_2]$   & 0.010 & 0.0071 & 0.0052 \\
		\hline
		Total  & 4.08 & 3.33 & 2.75 \\
		\hline
	\end{tabular}
	\caption{Decay widths (in unit: keV) for the production of $\Xi_{bc}$ via $W^+$ decays by varying $m_b$ (in unit: GeV).}
	\label{cbunb}
\end{table}

It seems that the mass deviation $m_c=1.8\pm0.3~\mathrm{GeV}$ brings larger influence to the width for $\Xi_{cc}$ production, than $m_b=5.1\pm0.3~\mathrm{GeV}$ to the width for $\Xi_{bc}$ production, but $0.3~\mathrm{GeV}$ is also relatively larger to $m_c$ than to $m_b$. We can change the mass of $c$ quark in $\Xi_{bc}$ baryon as well, however, it causes smaller difference than changing $m_b$. By varying $m_c=1.8\pm0.3~\mathrm{GeV}$, the total uncertainty for $\Xi_{bc}$ baryon is
\begin{eqnarray*}
	\Gamma(\Xi_{bc})=3.33^{+0.07}_{-0.03}~\mathrm{keV}.
\end{eqnarray*}
If adding these two uncertainties for $\Xi_{bc}$ baryon caused by $m_c$ and $m_b$ in quadrature, we can eventually obtain the total widths for $\Xi_{cc}$ and $\Xi_{bc}$ as follows
\begin{eqnarray}
	\Gamma(\Xi_{cc})&=&25.1^{+20.0}_{-9.8}~\mathrm{keV},\nonumber \\
	\Gamma(\Xi_{bc})&=&3.33^{+0.75}_{-0.58}~\mathrm{keV}.
\end{eqnarray}

\section{Summary}

In this paper, we investigate the production of the doubly heavy baryons $\Xi_{cc}$ and $\Xi_{bc}$ from the decay $W^+ \to \Xi_{Qc}+ \bar{Q}+\bar{s}$ under the NRQCD framework. High excited states are studied as well, include $[^1P_1]$, $[^3P_0]$, $[^3P_1]$, and $[^3P_2]$. Color antitriplet and sextuplet states are supposed to be of importance at the same level. The widths for different states and the total width are both well presented for convenience to see; the differential widths of angles and invariant masses are also given to promote our realization about the characteristics of these processes. The shapes of these figures are relevant to kinematics of the decays, and independent of those overall factors. Finally, the uncertainties are discussed by varying masses of the heavy constituent quarks $m_c=1.8\pm0.3~\mathrm{GeV}$ and $m_b=5.1\pm0.3~\mathrm{GeV}$.

Numerical results show that the contribution from the P-wave is generally one order less than the S-wave; $\Xi_{bc}$ baryon production is also about one order lower than $\Xi_{cc}$ production. For $\Xi_{cc}$, the total width is $25.1^{+20.0}_{-9.8}~\mathrm{keV}$, in which the P-wave occupies $6.3\%$. For $\Xi_{bc}$, the total width is $3.33^{+0.75}_{-0.58}~\mathrm{keV}$, in which the P-wave occupies $7.7\%$. These excited states may directly or indirectly (cascade-decay) decay to the ground states with a probability of almost $100\%$ via electromagnetic or hadronic interactions, so they are additional sources for the observed ground-state doubly heavy baryons in experiment. At the LHC running with the luminosity ${\cal L}=10^{34}~\rm{cm^{-2}\cdot s^{-1}}$, we can expect that there are totally $3.69\times10^5$ $\Xi_{cc}$ events and $4.91\times10^4$ $\Xi_{bc}$ events per operation year. In the paper, we do the integral in the whole phase space with no rapidity or transverse momentum cut. Thus, the yields above are the total events actually produced at the LHC, though there might be some events that are not recorded by the detectors in experiments for some inevitable reasons, like detectabilities.

Last of all, we discuss different colliders and the feasibility of experimental identifications about the production channels of doubly heavy baryons. There are two things that people are concerned about. Some of them focus on the final particles (like baryons here) properties with little interest in which process generates the particles, while others of them want to distinguish and study different processes. Generally speaking, we can reconstruct the decay processes from $W^+$ bosons to doubly heavy baryons, just as we reconstruct the doubly heavy baryons through their decay modes (like $\Xi_{c c}^{++} \rightarrow \Lambda_{c}^{+} K^{-} \pi^{+} \pi^{+}$, $\Lambda_{c}^{+} \rightarrow p K^{-} \pi^{+}$, and $\Xi_{bc}^{+} \rightarrow \Xi_{cc}^{++}\pi^{-}$). However, products of final states at hadron colliders are complex and bring many difficulties to our reconstruction. Besides, the dominant mechanism at the LHC is gluon-gluon fusion, and its yields are several magnitudes larger than yields of $W^+$ boson decays according to Refs.\cite{Chang:2006eu, Zhang:2011hi}. If one wants to deeply explore the production process, the better choice is $e^+ e^-$ colliders, on which we can precisely detect the heavy particles as well as the light $\bar{s}$ antiquark in final states under the clean background. This means that the reconstruction of $W^+$ boson decays into doubly heavy baryons is feasible on $e^+ e^-$ colliders, and some proposed Higgs factories are exactly capable for this target, such as the Future Circular Collider, the International Linear Collider, the Circular Electron Positron Collider, etc. Since the theoretical calculation on this indirect production is independent of colliders, our results about decay widths can be simply applied to these factories.

\hspace{2cm}

{\bf Acknowledgments}: This work was supported by the Natural Science Foundation of China under Grants No. 12005028 and No. 12047564, the Fundamental Research Funds for the Central Universities under Grant No. 2020CDJQY-Z003, and the China Postdoctoral Science Foundation under Grant No. 2021M693743.

\appendix

\section{The spin projectors and their derivatives for diquark $\bar{Q}c$}

The spin projectors and their derivatives are extensively used to calculate the hadron production. We have related the baryon production with the meson's. Now we give the relevant formulas; these formulas can be seen in Ref.\cite{Chang:2007si} but no detailed demonstration there.

Let $q=0$, with $\{\gamma^{\mu},\gamma^{\nu}\}=2g^{\mu\nu}$, we have
\begin{widetext}
\begin{eqnarray}\label{SPk11k12}
	\slashed{k}_{11}\slashed{k}_{12}&=&\frac{m_cm_Q}{M_{Qc}^2}\gamma^{\mu}\gamma^{\nu}(k_{1})_{\mu}(k_{1})_{\nu}\nonumber \\
	&=&\frac{m_cm_Q}{M_{Qc}^2}\frac{[\gamma^{\mu}\gamma^{\nu}(k_{1})_{\mu}(k_{1})_{\nu}+\gamma^{\nu}\gamma^{\mu}(k_{1})_{\nu}(k_{1})_{\mu}]}{2}\nonumber \\
	&=&\frac{m_cm_Q}{M_{Qc}^2}\frac{\gamma^{\mu}\gamma^{\nu}+\gamma^{\nu}\gamma^{\mu}}{2}(k_{1})_{\mu}(k_{1})_{\nu}\nonumber \\
	&=&\frac{m_cm_Q}{M_{Qc}^2}k_1^2=m_cm_Q.
\end{eqnarray}	
\end{widetext}

Using Eq.(\ref{SPk11k12}) and the properties $k_1^{\beta}\varepsilon^s_{\beta}(k_1)=0$ ($\varepsilon^s_{\beta}(k_1)$ is the polarization vector for S-wave mesons), we can simplify Eqs.(\ref{spinprojector1}, \ref{spinprojector3}) as
\begin{eqnarray}\label{simplify1}
	\Pi_{k_{1}}(0)&=&\frac{-\sqrt{M_{Qc}}}{4 m_{Q} m_{c}}\left(\slashed{k}_{12}-m_{Q}\right) \gamma^{5}\left(\slashed{k}_{11}+m_{c}\right)\nonumber \\
	&=&\frac{\sqrt{M_{Qc}}}{4 m_{Q} m_{c}}\gamma^{5}\left(\slashed{k}_{12}+m_{Q}\right)\left(\slashed{k}_{11}+m_{c}\right)\nonumber \\
	&=&\frac{\sqrt{M_{Qc}}}{4 m_{Q} m_{c}}\gamma^{5}\left(2m_cm_Q+2\frac{m_cm_Q}{M_{Qc}}\slashed{k}_1\right)\nonumber \\
	&=&\frac{1}{2\sqrt{M_{Qc}}}\gamma^{5}\left(\slashed{k}_1+M_{Qc}\right),
\end{eqnarray}
\begin{eqnarray}\label{simplify2}
	\Pi_{k_{1}}^{\beta}(0)&=&\frac{-\sqrt{M_{Qc}}}{4 m_{Q} m_{c}}\left(\slashed{k}_{12}-m_{Q}\right) \gamma^{\beta}\left(\slashed{k}_{11}+m_{c}\right)\nonumber \\
	&=&\frac{\sqrt{M_{Qc}}}{4 m_{Q} m_{c}}\left(\gamma^{\beta}\slashed{k}_{12}-2k_{12}^{\beta}+\gamma^{\beta}m_{Q}\right)\left(\slashed{k}_{11}+m_{c}\right)\nonumber \\
	&=&\frac{\sqrt{M_{Qc}}}{4 m_{Q} m_{c}}\gamma^{\beta}\left(\slashed{k}_{12}+m_{Q}\right)\left(\slashed{k}_{11}+m_{c}\right)\nonumber \\
	&=&\frac{1}{2\sqrt{M_{Qc}}}\gamma^{\beta}\left(\slashed{k}_1+M_{Qc}\right).
\end{eqnarray}

For P-wave amplitudes, the derivatives of Eqs.(\ref{spinprojector1}, \ref{spinprojector3}) are
\begin{widetext}
\begin{eqnarray}\label{simplify3}
	\left.\frac{d}{d q_{\alpha}} \Pi_{k_{1}}(q)\right|_{q=0}&=&\left.\frac{d}{d q_{\alpha}} \frac{-\sqrt{M_{Qc}}}{4 m_{Q} m_{c}}\left(\slashed{k}_{12}-m_{Q}\right) \gamma^{5}\left(\slashed{k}_{11}+m_{c}\right)\right|_{q=0}\nonumber \\
	&=&\left.\frac{\sqrt{M_{Qc}}}{4 m_{Q} m_{c}}[\gamma^{\alpha}\gamma^5\left(\slashed{k}_{11}+m_{c}\right)-\left(\slashed{k}_{12}-m_{Q}\right)\gamma^5\gamma^{\alpha}]\right|_{q=0}\nonumber \\
	&=&\left.\frac{\sqrt{M_{Qc}}}{4 m_{Q} m_{c}}[\gamma^{\alpha}\gamma^5\left(\slashed{k}_{1}+m_{c}-m_{Q}\right)-2\gamma^5(k_{12})^{\alpha}]\right|_{q=0}\nonumber \\
	&=&\frac{\sqrt{M_{Qc}}}{4 m_{Q} m_{c}}\gamma^{\alpha}\gamma^5\left(\slashed{k}_{1}+m_{c}-m_{Q}\right),
	\\
	\label{simplify4}
	\left.\frac{d}{d q_{\alpha}} \Pi^{\beta}_{k_{1}}(q)\right|_{q=0}&=&\left.\frac{d}{d q_{\alpha}} \frac{-\sqrt{M_{Qc}}}{4 m_{Q} m_{c}}\left(\slashed{k}_{12}-m_{Q}\right) \gamma^{\beta}\left(\slashed{k}_{11}+m_{c}\right)\right|_{q=0}\nonumber \\
	&=&\left.\frac{\sqrt{M_{Qc}}}{4 m_{Q} m_{c}}[\gamma^{\alpha}\gamma^{\beta}\left(\slashed{k}_{11}+m_{c}\right)-\left(\slashed{k}_{12}-m_{Q}\right)\gamma^{\beta}\gamma^{\alpha}]\right|_{q=0}\nonumber \\
	&=&\left.\frac{\sqrt{M_{Qc}}}{4 m_{Q} m_{c}}[\gamma^{\alpha}\gamma^{\beta}\left(\slashed{k}_{1}+m_{c}-m_{Q}\right)-2 g^{\alpha \beta}\left(\slashed{k}_{12}-m_{Q}\right)-2\gamma^{\alpha}k_{12}^{\beta}-2\gamma^{\beta}k_{12}^{\alpha}]\right|_{q=0}\nonumber \\
	&=&\left.\frac{\sqrt{M_{Qc}}}{4 m_{Q} m_{c}}[\gamma^{\alpha}\gamma^{\beta}\left(\slashed{k}_{1}+m_{c}-m_{Q}\right)-2 g^{\alpha \beta}\left(\slashed{k}_{12}-m_{Q}\right)]\right|_{q=0}.
\end{eqnarray}
\end{widetext}
Here $k_1^{\alpha}\varepsilon^l_{\alpha}(k_1)=0$ and $k_1^{\alpha}\varepsilon^J_{\alpha\beta}(k_1)=k_1^{\beta}\varepsilon^J_{\alpha\beta}(k_1)=0$ are uesed for the amplitudes of $[^1P_1]$ and $[^3P_J]$ states respectively.
\\

\end{document}